\documentclass[conference]{IEEEtran}
\IEEEoverridecommandlockouts
\usepackage{cite}
\usepackage{amsmath,amssymb,amsfonts}
\usepackage{graphicx}
\usepackage{textcomp}
\usepackage{xcolor}
\usepackage{hyperref}
\hypersetup{colorlinks,linkcolor={blue},citecolor={blue},urlcolor={blue}} 
\usepackage{url}

\hypersetup{nolinks=true}
\usepackage{ulem}
\usepackage{amssymb}
\usepackage{arydshln}
\usepackage{amsmath}
\usepackage{amsthm}
\usepackage[none]{hyphenat}
\usepackage{array}
\usepackage{enumitem}
\usepackage{tabularx}
\usepackage{longtable}
\usepackage{supertabular}
\usepackage{ltablex}
\usepackage{comment}
\usepackage{algpseudocode}
\usepackage{enumitem}
\usepackage[ruled,vlined]{algorithm2e}

\usepackage{booktabs}
\usepackage{flushend}
\def\BibTeX{{\rm B\kern-.05em{\sc i\kern-.025em b}\kern-.08em
    T\kern-.1667em\lower.7ex\hbox{E}\kern-.125emX}}
\begin{document}

\title{Unveiling Privacy Policy Complexity: An Exploratory Study Using Graph Mining, Machine Learning, and Natural Language Processing}

\author{\IEEEauthorblockN{1\textsuperscript{st} Vijayalakshmi Ramasamy*\thanks{*Corresponding author}}
\IEEEauthorblockA{\textit{Computer Science} \\
\textit{Georgia Southern University}\\
Statesboro, USA \\
vramasamy@georgiasouthern.edu}
\and
\IEEEauthorblockN{2\textsuperscript{nd} Seth Barrett}
\IEEEauthorblockA{\textit{Computer and Cybersciences} \\
\textit{Augusta University}\\
Augusta, USA \\
sebarrett@augusta.edu }
\and
\IEEEauthorblockN{3\textsuperscript{rd} Gokila Dorai}
\IEEEauthorblockA{\textit{Computer and Cybersciences} \\
\textit{Augusta University}\\
Augusta, USA \\
GDORAI@augusta.edu}
\and
\IEEEauthorblockN{4\textsuperscript{th} Jessica Zumbach}
\IEEEauthorblockA{\textit{Computer Science} \\
\textit{Georgia Southern University}\\
Statesboro, USA \\
jz00709@georgiasouthern.edu}
}
\maketitle
\begin{abstract}
Privacy policy documents are often lengthy, complex, and difficult for non-expert users to interpret, leading to a lack of transparency regarding the collection, processing, and sharing of personal data. As concerns over online privacy grow, it is essential to develop automated tools capable of analyzing privacy policies and identifying potential risks. In this study, we explore the potential of interactive graph visualizations to enhance user understanding of privacy policies by representing policy terms as structured graph models. This approach makes complex relationships more accessible and enables users to make informed decisions about their personal data (RQ1). We also employ graph mining algorithms to identify key themes, such as User Activity and Device Information, using dimensionality reduction techniques like t-SNE and PCA to assess clustering effectiveness. 
Our findings reveal that graph-based clustering improves policy content interpretability. It highlights patterns in user tracking and data sharing, which supports forensic investigations and identifies regulatory non-compliance. 
This research advances AI-driven tools for auditing privacy policies by integrating interactive visualizations with graph mining. Enhanced transparency fosters accountability and trust.
\end{abstract}


\section{Introduction}
\label{sec1}
In today’s fast-changing world, technology-driven applications are essential for addressing challenges and enhancing the quality of life. Mobile apps span various sectors, from healthcare tools like Fitbit and MyFitnessPal to communication platforms such as WhatsApp and Zoom. Financial apps like Google Pay and Robinhood simplify money management, while educational platforms like Coursera and Duolingo make learning accessible. Transportation apps like Uber and Google Maps improve mobility, and smart home solutions like Alexa and Nest enhance convenience and security.
These applications are vital for navigating modern life, promoting accessibility, productivity, and sustainability while also raising concerns about personal data security in cyberspace.

The growing use of mobile, healthcare, shopping, and gaming applications has intensified concerns about data privacy, as each app requires users to agree to privacy policies (PPs)—legal documents detailing how personal data is collected, used, and shared. While PPs serve as critical legal documents that inform users about how their personal data is collected, used, and shared by online platforms, many users accept these PPs without fully understanding them, creating risks of data misuse and loss of control over personal information. With apps managing sensitive data across platforms, enhanced safeguards, transparent practices, and simplified policies are crucial to protect users from breaches and cyber threats and diminish trust in digital services. 
Since PPs are written in complex legal terminology and tend to be lengthy, it is often unrealistic to expect users to thoroughly review and fully comprehend these detailed documents \cite{McDonald2009TheCO}. 
Consequently, this ambiguity raises significant concerns regarding the protection of consumer rights and the ethical obligations of service providers to maintain transparency and accountability~\cite{schairer2018commercial}.  A 2023 study by the Pew Research Center reveals that 79\% of adults are highly or somewhat concerned about how their personal information is used by companies. \cite{pew2019privacy}. Despite regulations like the General Data Protection Regulation (GDPR) and the California Consumer Privacy Act (CCPA), there remains a critical need for automated tools to help various stakeholders (consumers, data controllers, data processors, third-party partners, etc.) understand and analyze PPs efficiently \cite{schaub2015}.

\textbf{Motivation: } The lack of transparency in PPs can have significant implications, particularly in civil investigations and litigations, where investigators and legal professionals must analyze these policies to determine compliance with data protection regulations and assess potential risks to user privacy.
For example, in private-sector investigations, the ability to efficiently analyze and visualize PPs can help forensic experts uncover disparities between stated policies and actual data-handling practices. Visualizing key elements such as data-sharing practices, third-party involvement, and compliance with regulations like the CCPA can provide actionable insights for legal professionals and regulatory bodies. 

To this end, mobile applications across finance, commerce, and social media sectors have introduced significant privacy and security challenges. Financial applications have become essential tools for managing personal finances. Yet, they leave behind valuable digital traces that could serve as critical evidence in forensic investigations related to financial fraud and disputes \cite{dorai2023mint}. Similarly, the rapid growth of mobile commerce has transformed shopping behaviors, leading to an increased collection of personal data by e-commerce platforms. Forensic analyses of popular shopping applications have revealed substantial user data residues, raising concerns about privacy risks and the implications of default privacy settings \cite{dorai2023mobile}. Furthermore, through influencer marketing and hashtag-based interactions, the intersection of social media and e-commerce has introduced new dimensions of privacy threats, including cyberstalking and unauthorized data tracking \cite{dorai2022hashtag}.

In recent years, there have been notable legal cases highlighting the consequences of inadequate or misleading privacy practices. For instance, in 2024, a lawsuit was filed against TikTok~\cite{tiktok2024} for allegedly violating the Children’s Online Privacy Protection Act by collecting data from users under 13 without parental consent. The complaint emphasized TikTok’s failure to implement effective age verification and parental consent mechanisms, leading to unauthorized data collection from minors.  
Similarly, in 2025, LinkedIn faced a lawsuit~\cite{linkedin2025} for allegedly sharing private messages with third parties to train AI models without obtaining user consent. The lawsuit claimed that LinkedIn introduced a privacy setting allowing users to opt in or out of data sharing but later updated its PP to include AI model training data usage without proper user notification.  

While effective to some extent, the traditional methods of PP  analysis rely on rule-based and natural language processing (NLP) techniques, which struggle to accurately capture the intricate relationships between policy clauses, data entities, and legal obligations \cite{harkous2018}.
To address this gap, graph-based PP analysis offers a novel framework to model PPs as networks, reveal hidden structures, identify ambiguous clauses, and facilitate comparative analysis across different policies \cite{cui2023poligraph}. By leveraging graph mining, clustering methods, and visualization techniques, researchers can extract meaningful insights from the entities and the relationships described in PPs and enhance transparency in data practices. This article explores the motivations for using graph clustering and dimensionality reduction (DR) techniques, highlighting their significant benefits in predictive process analysis.

\textbf{Problem Statement and Research Questions: }
We present an empirical study analyzing Privacy Policies (PPs) from popular e-commerce mobile applications with large user bases. We employ DR techniques on knowledge graph (KG) data to simplify its complex structure and extract key insights from PPs, emphasizing data-sharing relationships, potential privacy risks, and compliance indicators.

Our research aims to connect intricate legal documentation with actionable insights for legal and regulatory stakeholders by providing interpretable visualizations for forensic investigators. The findings address the need for privacy-conscious software solutions that empower users and aid investigative efforts in private-sector cases and civil litigations. Additionally, the clustering techniques reveal hidden risks and enhance transparency through visual representation.
With these research objectives and goals, we propose the following overarching research questions:
\begin{itemize}[leftmargin=*]
\item \textbf{RQ1:} How can interactive graph visualizations of privacy policies, generated using Python libraries like NetworkX, Dash Cytoscape, and Plotly, enhance transparency and facilitate easier comprehension of complex policy terms for non-expert users, ultimately empowering them to make informed decisions about their personal data?
\item \textbf{RQ2:} How can graph mining algorithms be applied to represent, visualize, and identify the key structural properties of the graph communities in privacy policies? 
\end{itemize}
The remainder of this paper is structured as follows: Section 2 provides an overview of the terminology used and related research work in the literature on representing, visualizing, and analyzing privacy policies. Section 3 outlines our methodology, which uses a knowledge graph representation of PPs to facilitate insightful analyses of transparency and compliance, structural analysis, and automated extraction of privacy-related information using clustering algorithms. Section 4 presents the findings and validation of the results from our analysis and discusses their implications for forensic investigations and regulatory compliance, followed by a conclusion and future work.


\section{Related Work}


Research has shown that many mobile apps, financial tools, healthcare apps, and smart home solutions have inadequate PPs, which can lead to data breaches and unauthorized data sharing~\cite{Dahmen2017, Korunovska2020, Alamri2022}.
The increasing complexity and volume of PPs pose significant challenges for regulatory compliance, forensic investigations, and ensuring transparency in data handling practices. Traditional methods of reviewing these documents are time-intensive, lack scalability, and often fail to provide clear insights into potential areas of non-compliance or malicious intent.
It is crucial to support data owners or application users in comprehending the intricacies of PPs, especially when they are required to consent to and sign these agreements. 

Among the various techniques for representing text, graph representation stands out for providing a clear visual overview of the relationships and dependencies within the text. A \textit{graph}, also referred to as a \textit{network}, is a mathematical model consisting of nodes (representing entities, such as terms or concepts) and edges (representing relationship links or connections between these entities) used to represent and analyze multi-modal data in many different domains. This approach emphasizes the interconnectedness of information, allowing users to explore how different elements relate to one another in an intuitive and interactive manner~\cite{ramasamy2018tpgraphminer}. 
Graph-based approaches offer an effective alternative to textual analysis by structuring PPs as interconnected networks. A PP can be represented as a graph, where $Nodes$ represent key entities (e.g., personal data types, third parties, user rights, obligations) and $Edges$ denote relationships between these entities (e.g., ``data shared with third party", ``user consents to processing", ``company retains data"). This structured representation allows researchers to apply graph mining techniques to uncover hidden relationships and identify inconsistencies within PPs. 

The following articles in the literature discuss graph mining algorithms related to PP analysis. Examples of keywords used in these studies include ``graph mining privacy policies,'' ``network analysis in privacy policies,'' and ``graph-based privacy policy analysis.''
Boi (2023) investigated the application of knowledge graphs in decentralized data collection systems for managing and enforcing privacy policies, improving transparency and trust \cite{boi2023using}.
Cui (2023) presented PoliGraph, a system that leverages knowledge graphs for automating PP analysis, focusing on compliance issues and contradictions detection \cite{cui2023poligraph}.
Tauqeer (2022) introduced an automated KG-based tool for GDPR contract compliance verification (CCV) that binds GDPR’s legal basis to data-sharing contracts, facilitating effective compliance management \cite{tauqeer2022}.
Elluri (2018) developed an integrated KG ontology to automate GDPR and Payment Card Industry Data Security Standard (PCI- DSS) compliance, enhancing rule representation and application in data practices to validate against major vendors' data policies and enhance data protection.\cite{elluri2018integrated}.

These studies show that graph-based techniques are effective for uncovering implicit relationships between data practices and identifying risks not clearly stated in privacy policies (PPs). Analyzing PPs with graph similarity metrics allows for comparative studies of compliance and the discovery of non-conforming policies. Additionally, clustering techniques help categorize policies by their privacy practices, enabling stakeholders to assess alignment with regulatory frameworks. Graph visualization tools further simplify the understanding of data-sharing practices and user rights for non-experts.

Graph representations complement other forms of simplification by transforming PPs into analyzable information networks. They enhance user comprehension by revealing structural insights that linear or tabular formats often fail to capture. When combined with other representation methods, graphs complete the toolkit for simplifying complex textual information, ensuring a more inclusive and effective approach to understanding critical details. 
Interactive graph visualizations offer a promising solution by transforming PP text into clear and concise visual representations using tools like NetworkX, Dash Cytoscape, and Plotly~\cite{hagberg2008exploring, dash_cytoscape, plotly2015plotly}. These visualizations help investigators identify potential violations and areas of non-compliance. Graph mining algorithms also analyze communities within PPs, revealing patterns that may indicate malicious activities. However, the use of these techniques in forensic investigations is still under-explored.
This research explores how graph visualizations, graph mining, and automated extraction methods can enhance the understanding of privacy policies in forensic and regulatory contexts.

The following section describes the methodology of graph representation, visualization of the PPs, and clustering techniques used in this research.

\section{Methodology}
Clustering is an unsupervised learning technique to group a set of data points (or objects) into subsets, known as clusters, based on their similarity or proximity. In the context of graphs or networks, clustering involves identifying subsets of nodes that are more densely connected to each other than to the rest of the graph \cite{Rokach2005}. 
Our research uses the Poligraph framework ~\cite{cui2023poligraph}, an advanced tool for analyzing relationships between entities in policy documents (PPs). We also utilize PoliGraph-ER, which employs natural language processing (NLP) techniques to extract relationships from the text. The implementation is based on the authors' source code from their publicly available repository~\cite{PoliGraphGitHub2023, PolicyTechnology2023}.

\textbf{Graph Representation: }The graphML files generated using the Poligraph-ER algorithm consist of $Nodes$ with the attributes Node ID, label, and type (e.g., Node ID, label: ``information you provide to we include, type: DATA), and $Edges$ with the attributes Source, Target, relationship, text, and id (e.g., Source: ``information you provide to we include", Target: ``information you register", relationship: ``SUBSUM", text=	``OfferUp collects information that you provide directly to us when you sign up and use the OfferUp service including Information when you register or update the details of your account,'' and	id = ``e0.'' 

The following \textbf{Dimentionality Reduction (DR) }  techniques, leveraged with the clustering techniques, were used to compare their performance in various clustering methods.
(i) \textbf{t-SNE} t-distributed Stochastic Neighbor Embedding: Non-linear dimensionality reduction that preserves local structure in high-dimensional data \cite{Nassar2024tSNE}.
(ii) \textbf{UMAP} Uniform Manifold Approximation and Projection: Non-linear dimensionality reduction that preserves global and local structure in low dimensions \cite{McInnes2018UMAP}. (iii) \textbf{PCA} Principal Component Analysis: Linear dimensionality reduction that projects data onto principal components of variance \cite{FRS1901LIIIOL}.

Below is an overview of the clustering techniques used in the article and their suitability for PP text data visualization.
\textbf{Mini-Batch K-means} (MB-K-Means) Alternative to K-means, used for clustering massive datasets to reduce computational cost ~\cite{sculley2010web}: Well-suited for large-scale PPs to efficiently group similar policy clauses and terms and detect common privacy themes across different documents. 
\textbf{DBSCAN} (Density-Based Spatial Clustering of Applications
with Noise) - Groups closely packed points and labels low-density points as outliers~\cite{Ester1996DBSCAN}: Useful to identify ambiguous terms or inconsistent privacy clauses in noisy data or when cluster shapes are irregular.
\textbf{HDBSCAN} (Hierarchical DBSCAN) - Extension of DBSCAN, automatically determines number of clusters, identifies noise~\cite{campello2013density}: Useful for identifying meaningful clusters and outliers in PP text.
\textbf{Spectral Clustering -} Graph-based, uses eigenvalues of similarity matrix (graph Laplacian) for dimensionality reduction before clustering~\cite{li2020privacy}: Best for PP clauses represented as similarity graphs (e.g., semantic similarity between terms or policies).
\textbf{LDA:} Latent Dirichlet Allocation - A probabilistic topic modeling method that clusters words into topics and assigns probabilities of topics to each document (soft clustering)~\cite{kuznetsov2021towards}: Useful to summarize PP text into themes (e.g., ``Data Sharing'', ``Consent''), helping identify key compliance areas or risks.

Using the PoliGraph-ER algorithm and the implementation provided by Cui et al.,~\cite{cui2023poligraph} described in the previous section, we constructed knowledge graphs for PPs sourced from Android e-commerce privacy policies. The resulting knowledge graphs were stored in standardized file formats, including graphyml and graphml, for efficient representation and analysis. The graphs stored as graphML represent complex nonlinear relationships between nodes and edges. 
The PoliGraph-ER framework primarily focuses on data collection practices. This emphasis may limit its flexibility in meeting basic user requirements for an app, such as providing a visual representation of the structure and various components of PPs. Therefore, we used a graph mining framework that could enhance its utility by representing, visualizing, and identifying key structural properties of graph communities (clusters) within these policies. 
The twofold analysis phases used for the visualization and clustering of PPs to gather structural and semantic insights are presented below:
\textit{Phase 1:} We first collected URLs that link to the PPs of 55 Android e-commerce apps. 
The PoliGraph-ER tool preprocesses the PP Text to remove irrelevant data and normalize the content, segment them into sentences, and extract entities (nodes) and relationships (edges) using advanced NLP techniques like named entity recognition (NER) and dependency parsing. These extracted elements are mapped onto a graph structure, where entities form nodes and relationships define the edges. 
\textit{Phase 2:} We implemented code using the Dash Cytoscape, an interactive and highly customizable graph visualization library, to create dynamic, visually rich, and web-based graph representations. 
The graph embeddings from the GraphML files were visualized using popular ``non-linear'' DR techniques t-SNE and UMAP, and the ``linear'' DR technique PCA. 

The steps for applying graph clustering algorithms, including node embedding dimensionality reduction, are shown in Algorithm 1.
\textit{Step 1:} Load the GraphML file into a directed graph G.
\textit{Step 2:} Construct the Directed Graph.
\textit{Step 3:} Convert the Spring layout positions (to compute node positions to minimize edge crossings and making the graph structure easier to interpret) into a high-dimensional embedding matrix.
\textit{Step 4:} Apply t-SNE, UMAP, and PCA to reduce the embedding matrices to 2D space for visualization.
\textit{Step 5:} Perform Clustering and evaluation of MB K-Means, Agglomerative, HDBSCAN (based on density, identifying clusters and noise points), Spectral Clustering (based on graph affinity), and LDA algorithms by applying the method to the reduced embeddings (t-SNE, UMAP, and PCA), validate results to assess clustering quality.
\textit{Step 6:} Visualize the results using scatterplots of clustering results for each method and t-SNE, UMAP, and PCA embeddings.
\begin{algorithm}[htbp]
\begin{small}
\caption{Graph Clustering and Visualization}
\KwIn{
    \textbf{}GraphML file containing:
    \begin{itemize}
        \item[ ] \textit{Edge Data}: source, target, relationship, and text
        \item[ ] \textit{Node Data}: node ID, label, type.
    \end{itemize}
}
\KwOut{Cluster Visualizations - t-SNE, UMAP, and PCA embeddings; Eval. metrics: Silhouette scores, Davies-Bouldin Index for MB K-Means, Agglom., HDBSCAN, Spectral Clustering, LDA
}
\textbf{Step 1: Load Data} \\
\texttt{edge\_data} $\gets$ Read \textit{Edge Data}  \\
\texttt{node\_data} $\gets$ Read \textit{Node Data}

\textbf{Step 2: Construct Graph} \\
$G \gets$ DirectedGraph() 

\ForEach{ row in \texttt{node\_data}}{
    Add node to $G$ with attr: \texttt{label}, \texttt{type}
}

\ForEach{ row in \texttt{edge\_data}}{
    Add directed edge to $G$ with attr: \texttt{relationship}, \texttt{text}
}

\textbf{Step 3: Compute Node Embeddings} \\
\texttt{pos} $\gets$ Spring layout positions for $G$ \\
\texttt{node\_embed} $\gets$ DataFrame of \texttt{pos} 

\textbf{Step 4: Perform Dimensionality Reduction} \\
\texttt{tsne\_embed} $\gets$ Apply t-SNE to \texttt{node\_embed} \\
\texttt{umap\_embed} $\gets$ Apply UMAP to \texttt{node\_embed} 

\textbf{Step 5: Evaluate Clustering} \\
\SetKwFunction{EvaluateAndVisualize}{EvaluateAndVisualize}
\EvaluateAndVisualize{embeddings, method\_name}\;
Evaluate and plot performance of each $method\_name$ using Silhouette Score and Davies-Bouldin Index\;

\textbf{Step 6: Visualize for t-SNE and UMAP} \\
\EvaluateAndVisualize(\texttt{tsne\_embeddings}, \texttt{"t-SNE"}) \\
\EvaluateAndVisualize(\texttt{umap\_embeddings}, \texttt{"UMAP"}) \\
\end{small}
\end{algorithm}

\textbf{Cluster Validation Techniques: }
When evaluating the clustering of PPs based on DR techniques like t-SNE, UMAP, and PCA, we used \textbf{\textit{Silhouette Score}} to determine how well the PP data points that are grouped together in clusters are internally cohesive and externally separated. 
The \textit{Silhouette Score} metric (range: -1 to +1) measures how similar an object $a$ is to its own cluster (cohesion) compared to other clusters (separation)
A high value (close to +1) indicates that the point is well-matched to its own cluster and poorly matched to neighboring clusters.
A value of 0 indicates the point is on or very close to the decision boundary between two neighboring clusters.
A negative value indicates that the point might have been assigned to the wrong cluster. 
The \textbf{\textit{Davies-Bouldin Index}} (DBI) is a measure of the clustering quality indicating the average ‘similarity’ between clusters, i.e., the average similarity measure of each cluster with its most similar cluster, where similarity is the ratio of within-cluster distances to between-cluster distances. 
Lower values of the Davies-Bouldin Index (DBI) indicate better clustering with minimal intra-cluster scatter, while higher values suggest poor clustering with overlaps. In the context of PP clustering, a low DBI signifies well-separated and compact clusters, ideal for analysis. This means that entities within each cluster share specific traits, whereas those in different clusters have distinct features. 
\section{Results and Discussion}
Our contribution to understanding privacy policies (PPs) through knowledge graphs is two-fold. First, we present the results of visualizing the Privacy Policy Knowledge Graph (PPKG), which was derived from a GraphML file. Additionally, we expanded our experiments by applying clustering techniques to gain deeper insights into how PPs can be viewed in a structured manner.

\begin{figure*}[hbt!]
\centering
\includegraphics[width=0.725\textwidth]{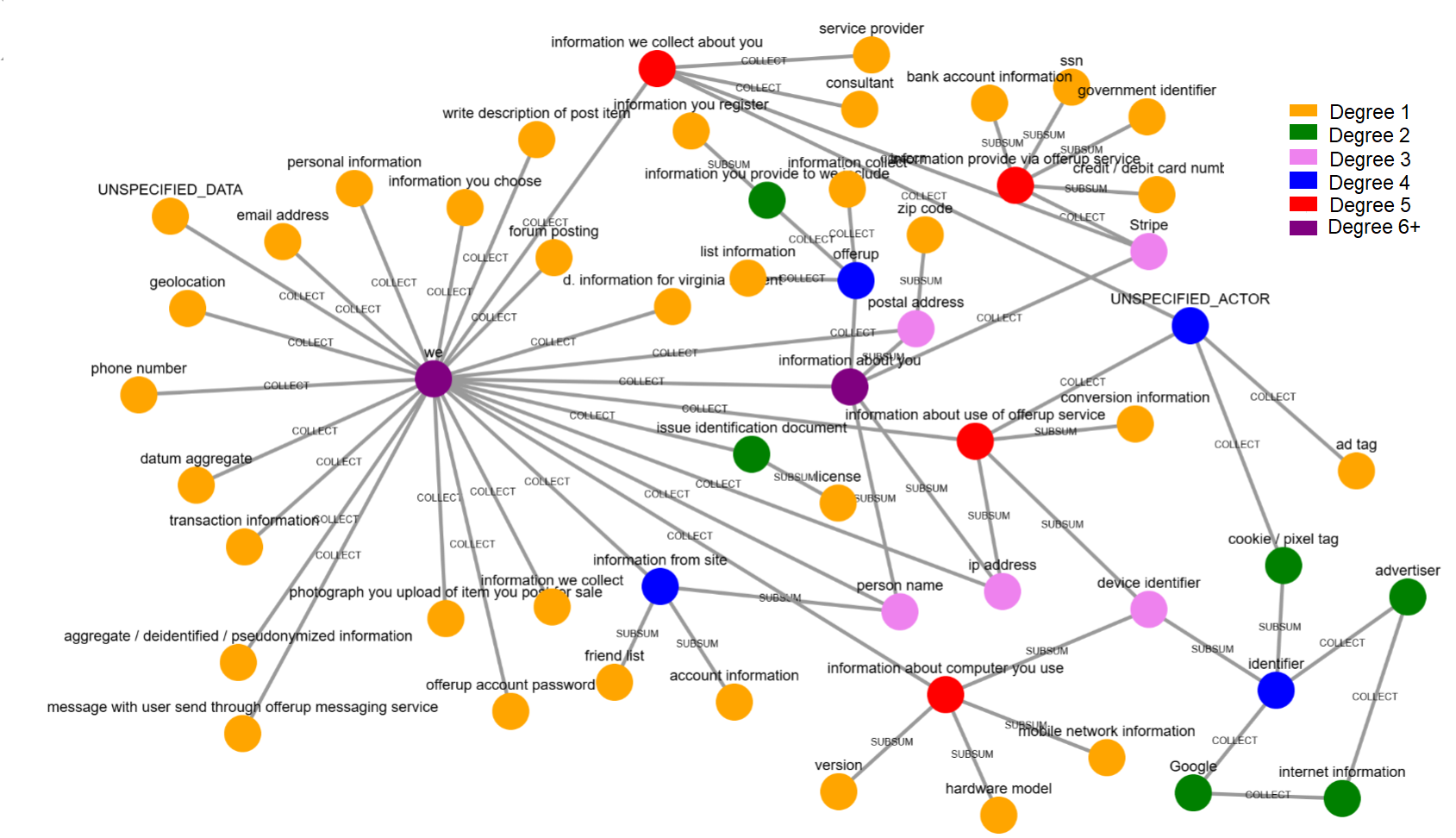}
\caption{Privacy Policy Graph Visualization of Data Collection and Interaction Example}\label{fig:offerup}
\end{figure*}

\textbf{Privacy Policy Knowledge Graph Visualization: }
First, the PPKG of Offerup~\cite{wingfield2015offerup} PP in Fig.~\ref{fig:offerup} illustrates a holistic view of how data elements (nodes) and their relationships (edges) are structured in a directed graph. 
Using a degree-based color scheme for nodes enhances the visualization, making understanding the cohesively connected data elements easier. The \textit{Dash Cytoscape} provides an interactive interface where the users can explore the PPs data and derive insights by moving the nodes. The edge annotation of the \textit{\textquotedblleft COLLECT"} relationship in PoliGraph signifies the action of collecting data, as explicitly outlined in PPs. This annotation is used to identify the entities, such as companies or service providers, that collect specific types of user data, including email addresses, geolocation, and phone numbers. 

The PPKG visualization offers insights into the collection of various types of data, such as email addresses, geolocation, and transaction information. It highlights the aggregation and processing of this collected data into aggregate forms with a focus on data minimization.
Thus, the PPKG provides a detailed overview of the complexity of data practices, including the types of data collected, how they are categorized, and the relationships between them. This information can be used to systematically compare PPs across organizations, identify trends and potential privacy risks, and assess compliance with regulations such as GDPR and CCPA.

\begin{figure*}[ht]
\centering
\includegraphics[width=0.9\textwidth]{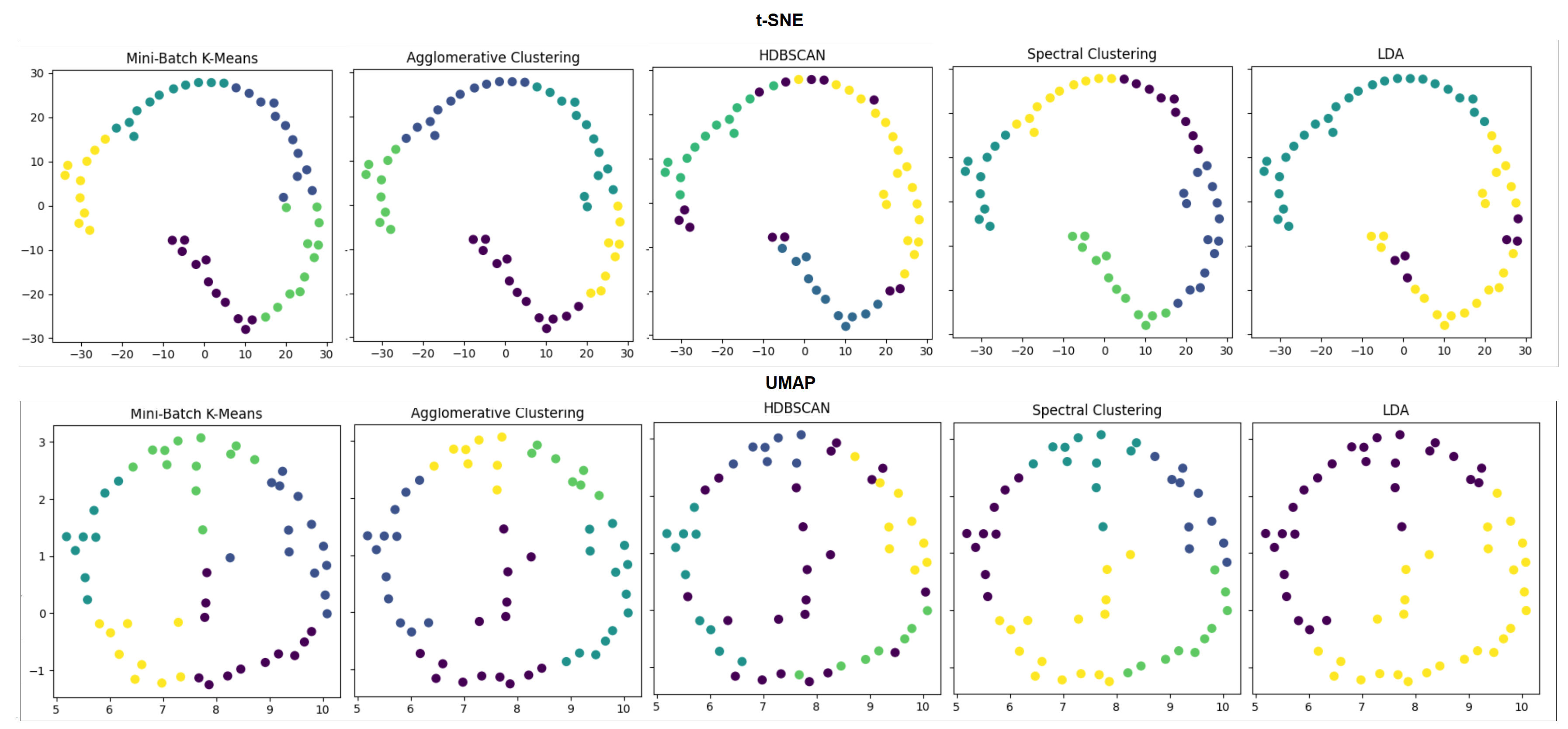}
\caption{Evaluating Privacy Policy Clustering Techniques using t-SNE and UMAP  }\label{fig:clustering}
\end{figure*}

\textbf{Clustering Validation: 
}
To explore the PPKGs further, we used a wide range of clustering techniques as shown in \textbf{\textit{Algorithm 1} }to group the data elements based on `User Activity,' `Device Information,' `Media and Location,' `Device and Advertising,' and `Unique Device Identification.' Fig.~\ref{fig:clustering} shows the results of applying various clustering techniques on t-SNE and UMAP embeddings. 
MB K-Means demonstrates a clear separation of clusters across all three embeddings: t-SNE, UMAP, and PCA (not shown due to space restrictions). Agglomerative Clustering performs similarly to K-Means but maintains stronger relationships between closely positioned points in the original data, effectively preserving local structure in all embeddings. Spectral Clustering exhibits smooth and well-defined transitions between clusters in both UMAP and PCA embeddings. In contrast, the results from LDA are less distinct than those from the other algorithms. This is likely due to LDA's probabilistic topic-modeling nature, which may not be optimal for this type of clustering task.
The performance evaluation of the various clustering algorithms are evaluated based on the Silhouette Scores and Davies-Bouldin Index (BDI) are shown in Table~\ref{tab:comparesilhouette} are as follows:
\begin{table}[h]
\centering
\caption{Performance Evaluation of Clustering Algorithms }\label{tab:comparesilhouette}
\begin{small}
\begin{tabular}{p{1.75cm} p{1.55cm} p{1.25cm} p{1.55cm} }
\toprule
\textbf{Clustering} & \textbf{t-SNE} & \textbf{UMAP} &\textbf{PCA} \\
\toprule
\multicolumn{4}{l}{Silhouette Scores}\\
\cdashline{1-4}
 MB K-means  &  0.4589 & 0.4119 &0.3712\\
 Agglomerative      & 0.4479   & 0.3856 &0.8033\\
 HDBSCAN            & 0.2907   & 0.1591 & 0.0944\\ 
 Spectral           &  0.4747  & 0.4115 & 0.3729\\
 LDA                &  0.1758  & 0.3539& 0.3333\\
 
\toprule
\multicolumn{4}{l}{Davies-Bouldin Index}\\
\cdashline{1-4}
 MB K-means  &  0.6623 & 0.7206 &0.7801\\
 Agglomerative      & 0.6318 & 0.7278& 0.8033\\
 HDBSCAN            & 1.5047 & 1.2096& 1.6524\\ 
 Spectral           &  0.6431 &0.6896& 0.7914\\
 LDA                &  3.8388& 1.1936 & 1.2114\\
 
\toprule

\end{tabular}
\end{small}
\end{table}

The results based on the performance evaluation of the various clustering algorithms are evaluated based on the Silhouette Scores and Davies-Bouldin Index shown in Table~\ref{tab:comparesilhouette} are as follows:
t-SNE generally has higher silhouette scores than UMAP across most clustering algorithms. PCA shows the lowest values. Similarly, t-SNE exhibits the lowest DBI, indicating better cluster separation, followed by UMAP and PCA. t-SNE shows the best performance in both silhouette scores and DBI for most algorithms, suggesting it creates better-separated clusters. It can be noted that t-SNE is slightly better suited for algorithms sensitive to cluster shapes, like Spectral Clustering and MB K-means. UMAP consistently achieves moderately high silhouette scores and DBI values, suggesting decent cluster separation. It performs comparably to t-SNE for algorithms like Spectral Clustering but struggles more with HDBSCAN and LDA. On the contrary, PCA generally shows the lowest silhouette scores and higher DBI, indicating poor cluster separation compared to t-SNE and UMAP.
Agglomerative clustering on PCA is an exception, showing relatively strong performance.

Overall, \textit{Min-Batch K-means } achieved the best silhouette score (t-SNE (0.4589), followed by UMAP (0.4119) and PCA (0.3712)) and BDI (t-SNE (0.6623), followed by UMAP (0.7206), and PCA (0.7801)). All the other clustering algorithms also show a similar trend. 
Since DBSCAN showed limited effectiveness in identifying only one or two clusters, we did not include the results. 

Based on the detailed evaluation, t-SNE and UMAP prove more effective for visualizing and clustering PP data compared to PCA, as they better capture the \textit{non-linear relationships} and complex structures within the data. Among the clustering algorithms, MB K-Means shows consistently moderate to good performance across different embeddings, which means it is a reliable choice for PP data analysis. Agglomerative clustering generally performs well, particularly with t-SNE and UMAP embeddings, indicating its suitability with the right parameter tuning. HDBSCAN and Spectral clustering underperform across all types of embeddings used in the study, indicating that these methods may not be useful for this specific dataset or require significant tuning to yield meaningful results. 

The findings in this article support Research Question 1 (RQ1) by establishing a foundation for transforming privacy policies into interactive and user-friendly visualizations. Additionally, they support Research Question 2 (RQ2) by utilizing graph mining algorithms to uncover the structural properties of semantic clusters, which allows for a deeper analysis of PPs. Together, these insights enhance transparency and help identify key themes within PP data.
Overall, this study lays the groundwork for integrating these methods into forensic procedures while also advancing the field of PP research. The suggested techniques provide investigators the opportunity to ensure accountability, defend user rights, and promote confidence in digital ecosystems by increasing openness and facilitating a methodical assessment of PPs.
\section{Conclusion and Future Work}


This study examines the use of graph-based clustering techniques to analyze PPs, focusing on user activity tracking, device information sharing, and third-party data disclosures. By applying dimensionality reduction methods such as t-SNE, UMAP, and PCA, different clustering approaches were evaluated for their effectiveness in identifying patterns within PPs. The visualizations and analyses demonstrate how clustering and graph-based techniques can simplify complex PPs by grouping related terms into semantically meaningful categories, such as user activity or device information (RQ1). These clusters, generated using algorithms like MB K-Means and HDBSCAN, highlight thematic communities and relationships between terms, providing structural insights into the policies (RQ2). 
The results suggest that these methods can support forensic experts and regulatory bodies in comparing policies, identifying variations, and assessing potential compliance issues.

Future research could refine this approach by integrating more advanced natural language processing techniques for improved entity recognition and context analysis. Expanding the dataset to include privacy policies from various industries may help assess the generalizability of these techniques. Additionally, developing automated tools that incorporate interactive visualizations could enhance the usability of PP analysis for both researchers and practitioners. These directions may contribute to improving transparency in PP interpretation and supporting investigations into data governance practices.

\Urlmuskip=0mu plus 1mu\relax
\bibliographystyle{elsarticle-num} 
\bibliography{references}

\begin{thebibliography}{10}
\expandafter\ifx\csname url\endcsname\relax
  \def\url#1{\texttt{#1}}\fi
\expandafter\ifx\csname urlprefix\endcsname\relax\def\urlprefix{URL }\fi
\expandafter\ifx\csname href\endcsname\relax
  \def\href#1#2{#2} \def\path#1{#1}\fi

\bibitem{McDonald2009TheCO}
A.~M. McDonald, L.~F. Cranor, The cost of reading privacy policies, 2009.

\bibitem{schairer2018commercial}
C.~E. Schairer, C.~K. Rubanovich, C.~S. Bloss, How could commercial terms of use and privacy policies undermine informed consent in the age of mobile health?, AMA Journal of Ethics 20~(9) (2018) E864--872.

\bibitem{pew2019privacy}
{Pew Research Center}, Americans and privacy: Concerned, confused and feeling lack of control over their personal information, accessed: 2025-01-30 (2019).

\bibitem{schaub2015}
F.~Schaub, R.~Balebako, A.~L. Durity, L.~F. Cranor, A design space for effective privacy notices, in: Proceedings of the Eleventh USENIX Conference on Usable Privacy and Security, SOUPS '15, USENIX Association, USA, 2015, p. 1–17.

\bibitem{dorai2023mint}
R.~Bardhan, R.~Garay-Paravisini, G.~Dorai, L.~Vanputte, Digital forensics analysis of a financial mobile application: Uncovering security and privacy implications, in: 2024 International Symposium on Networks, Computers and Communications (ISNCC), 2024, pp. 1--8.

\bibitem{dorai2023mobile}
G.~Dorai, S.~Hutchinson, Mobile commerce-analysis and investigation of the online safety, privacy, and data forensics of amazon and etsy apps, in: Proceedings of the 56th Hawaii International Conference on System Sciences, 2023.

\bibitem{dorai2022hashtag}
B.~Sumner, G.~Dorai, J.~Heslen, Preliminary analysis of privacy implications observed in social-media posts across shopping platforms, in: Proceedings of the 17th International Conference on Availability, Reliability and Security, ARES '22, Association for Computing Machinery, New York, NY, USA, 2022.

\bibitem{tiktok2024}
B.~Makuch, U.s. sues tiktok for collecting kids' data without parents' permission, The VergeAccessed: 2025-01-25 (2024).

\bibitem{linkedin2025}
D.~Levine, Microsoft's linkedin sued for disclosing customer information to train ai models, ReutersAccessed: 2025-01-25 (2025).

\bibitem{harkous2018}
H.~Harkous, K.~Fawaz, R.~Lebret, F.~Schaub, K.~G. Shin, K.~Aberer, Polisis: automated analysis and presentation of privacy policies using deep learning, in: Proceedings of the 27th USENIX Conference on Security Symposium, SEC'18, USENIX Association, USA, 2018, p. 531–548.

\bibitem{cui2023poligraph}
H.~Cui, R.~Trimananda, A.~Markopoulou, S.~Jordan, Poligraph: Automated privacy policy analysis using knowledge graphs, in: 32nd USENIX Security Symposium, 2023.

\bibitem{Dahmen2017}
J.~Dahmen, D.~J. Cook, X.~Wang, W.~Honglei, Smart secure homes: A survey of smart home technologies that sense, assess, and respond to security threats, Journal of Reliable Intelligent Environments 3~(2) (2017) 83--98.

\bibitem{Korunovska2020}
J.~Korunovska, B.~Kamleitner, S.~Spiekermann, The challenges and impact of privacy policy comprehension (Jun. 2020).

\bibitem{Alamri2022}
H.~Alamri, C.~Maple, S.~Mohamad, G.~Epiphaniou, Do the right thing: A privacy policy adherence analysis of over two million apps in apple ios app store, Sensors 22~(22) (2022) 8964.

\bibitem{ramasamy2018tpgraphminer}
V.~Ramasamy, U.~Desai, H.~W. Alomari, J.~D. Kiper, Tp-graphminer: A clustering framework for task-based information networks, in: 2018 IEEE International Conference on System, Computation, Automation and Networking (ICSCA), IEEE, 2018, pp. 1--6.

\bibitem{boi2023using}
B.~Boi, C.~Esposito, Using knowledge graphs to ensure privacy policies in decentralized data collection systems, in: International Conference on Research in Adaptive and Convergent Systems, ACM, 2023, p.~6.

\bibitem{tauqeer2022}
A.~Tauqeer, A.~Kurteva, T.~R. Chhetri, A.~Ahmeti, A.~Fensel, Automated gdpr contract compliance verification using knowledge graphs, Information 13~(10) (2022).

\bibitem{elluri2018integrated}
L.~Elluri, A.~Nagar, K.~P. Joshi, An integrated knowledge graph to automate gdpr and pci dss compliance, in: Proceedings of the 2018 IEEE International Conference on Big Data, University of Maryland Baltimore County, IEEE, Baltimore, MD, USA, 2018.

\bibitem{hagberg2008exploring}
A.~A. Hagberg, P.~J. Swart, D.~A. S~Chult, Exploring network structure, dynamics, and function using networkx, Proceedings of the 7th Python in Science Conference (SciPy2008) 2008 (2008) 11--15.

\bibitem{dash_cytoscape}
P.~T. Inc., Dash cytoscape: A component library for interactive graph visualization in python (2021).

\bibitem{plotly2015plotly}
P.~T. Inc., Plotly: Collaborative data science, Plotly (2015).

\bibitem{Rokach2005}
L.~Rokach, O.~Maimon, Clustering Methods, Springer US, Boston, MA, 2005, pp. 321--352.

\bibitem{PoliGraphGitHub2023}
{UCI Networking Group}, Poligraph github repository (including the sourcecode), accessed: January 2025 (June 2023).

\bibitem{PolicyTechnology2023}
{UCI Networking Group}, Policy-technology | uci networking group, accessed: January 2025 (June 2023).

\bibitem{Nassar2024tSNE}
S.~H. Nassar, S.~Belhaouari, M.~Allaoui, M.~L. Kherfi, Centroid initialization method for t-distributed stochastic neighbour embedding (t-sne), in: Proceedings of the 2024 IEEE International Conference on Machine Learning and Natural Language Processing (MLNLP), IEEE, 2024, pp. 1--11.

\bibitem{McInnes2018UMAP}
L.~McInnes, J.~Healy, J.~Melville, {UMAP}: Uniform manifold approximation and projection for dimension reduction (2018).
\newblock \href {http://arxiv.org/abs/1802.03426} {\path{arXiv:1802.03426}}.

\bibitem{FRS1901LIIIOL}
K.~P. F.R.S., Liii. on lines and planes of closest fit to systems of points in space, Philosophical Magazine Series 1 2 (1901) 559--572.

\bibitem{sculley2010web}
D.~Sculley, Web-scale k-means clustering, in: Proceedings of the 19th International Conference on World Wide Web (WWW'10), Association for Computing Machinery, 2010, pp. 1177--1178.

\bibitem{Ester1996DBSCAN}
M.~Ester, H.-P. Kriegel, J.~Sander, X.~Xu, A density-based algorithm for discovering clusters in large spatial databases with noise, in: Proceedings of the 2nd International Conference on Knowledge Discovery and Data Mining (KDD'96), AAAI Press, Portland, OR, United States, 1996, pp. 226--231.

\bibitem{campello2013density}
R.~J. G.~B. Campello, D.~Moulavi, J.~Sander, Density-based clustering based on hierarchical density estimates, in: J.~Pei, V.~S. Tseng, L.~Cao, H.~Motoda, G.~Xu (Eds.), Advances in Knowledge Discovery and Data Mining: 17th Pacific-Asia Conference, PAKDD 2013, Proceedings, Part II, Springer, 2013, pp. 160--172.

\bibitem{li2020privacy}
J.~Li, J.~Wei, M.~Ye, W.~Liu, X.~Hu, Privacy-preserving constrained spectral clustering algorithm for large-scale data sets, IET Information Security 14~(2) (2020) 175--184.

\bibitem{kuznetsov2021towards}
M.~Kuznetsov, E.~Novikova, Towards application of text mining techniques to the analysis of the privacy policies, in: Proceedings of the 10th Mediterranean Conference on Embedded Computing (MECO), IEEE, 2021.

\bibitem{wingfield2015offerup}
N.~Wingfield, Offerup takes on craigslist with war chest and mobile strategy, Bits Blog, archived from the original on 2020-01-24. Retrieved 2020-03-26 (Nov. 2015).

\end{thebibliography}



\end{document}